\documentclass[twocolumn,prl,aps,showpacs,floatfix,superscriptaddress]{revtex4}
\usepackage{dcolumn}
\usepackage{bm}

\usepackage{amsmath}
\usepackage{amsfonts}
\usepackage{graphics}
\usepackage{psfrag}
\usepackage{graphicx}
\usepackage{epsfig}
\usepackage{rotating}
\usepackage[latin1]{inputenc}
\usepackage{color}
\usepackage{rotating}

\begin{document}
\title{Helix or Coil?  Fate of a Melting Heteropolymer}
\author{G. Oshanin}
\affiliation{Laboratoire de Physique Th\'eorique de la Mati\`ere Condens\'ee,
  Universit\'e Pierre et Marie Curie, 4 Place Jussieu, 75252 Paris, France}
\author{S. Redner}
\affiliation{Laboratoire de Physique Th\'eorique de la Mati\`ere Condens\'ee,
  Universit\'e Pierre et Marie Curie, 4 Place Jussieu, 75252 Paris, France}
\affiliation{Center for Polymer Studies and Department of Physics, Boston
  University, Boston, Massachusetts 02215 USA} \date{\today}

\begin{abstract}
  We determine the probability that a partially melted heteropolymer at the
  melting temperature will either melt completely or return to a helix state.
  This system is equivalent to the splitting probability for a diffusing
  particle on a finite interval that moves according to the Sinai model.
  When the initial fraction of melted polymer is $f$, the melting probability
  fluctuates between different realizations of monomer sequences on the
  polymer.  For a fixed value of $f$, the melting probability distribution
  changes from unimodal to a bimodal as the strength of the disorder is
  increased.

\end{abstract}

\maketitle

\begin{figure}[ht]
  \centerline{\includegraphics*[width=0.285\textwidth]{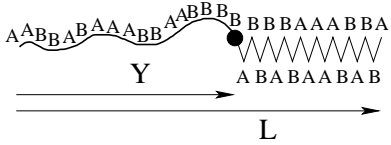}}
\caption{Illustration of a partially melted heteropolymer with the
  interface between the melt (left) and helix (right) at $Y$.}
  \label{cartoon}
\end{figure}

Thermally-induced helix-coil transitions in biological heteropolymers exhibit
intriguing thermodynamic and kinetic features \cite{melt1,melt2,melt3,melt4}
that stem from, for example, different intrinsic melting temperatures of the
monomeric constituents, say, $A$ and $B$, and {\em quenched\/} random
distributions of monomers along the chain.  Due to this randomness, an
arbitrarily long $A$-rich helix region with a high local melting temperature
can act as a barrier to hinder the melting of the entire chain into a random
coil.  As the temperature is raised, the nucleation of multiple coils may
occur at distinct points along the condensed helical chain.  Nevertheless, as
considered in Ref.~\cite{deG}, the simplified situation in which there is a
single coil portion starting at a free end of the chain (Fig.~\ref{cartoon})
highlights the essential physical role of randomness on melting kinetics.

Our primary result is that the probability that a partially melted
heteropolymer returns to its native helical state (and the complementary
probability that the chain denatures) exhibits huge sample-specific
fluctuations.  Consequently, the average and the typical behaviors are
completely different, and neither is representative of the behavior of a
single chain.  Our analysis is based on making an analogy \cite{deG} between
the motion of the boundary point (Fig.~\ref{cartoon}) in a partially melted
heteropolymer and a particle diffusing in the presence of random force---the
Sinai model \cite{sinai}.  

Let $E_+(Y)$ be the melting probability for a heteropolymer of (contour)
length $L$ with a partially melted segment of length $Y$, while
$E_-(Y)=1-E_+(Y)$ is the probability that this heteropolymer condenses.  The
melting probability is equivalent to the {\em splitting probability} $E_+(Y)$
that the boundary point starts at $Y$ and reaches $L$ without ever reaching
$0$ (and correspondingly for $E_-(Y)$).  For a random potential that
corresponds to a heteropolymer, the splitting probabilities are different for
each sequence of monomers.  Moreover, the resulting {\em distribution} of
splitting probabilities changes from a single peak at its most probable value
for weak disorder (with a delta-function peak in the absence of disorder), to
double peaked with most probable values close to $0$ and $1$ for strong
disorder.  Thus much care is needed to interpret experimental data of
heteropolymer melting kinetics.

Although the monomer sequence in a heteropolymer is typically correlated over
a finite range, we make the simplifying assumption that such correlations are
absent.  The location of the equivalent boundary point between the coil and
helix portions of the chain then diffuses in the interval $[0,L]$ in presence
of a quenched and random position-dependent force $F(x)$, with
\begin{align}
\label{force}
&\langle F(x) \rangle = F_0, \nonumber\\
&\langle F(x) F(x') \rangle - F_0^2 = \sigma^2_F \delta(x - x').
\end{align}
The mean force $F_0>0$ when the temperature $T$ exceeds the heteropolymer
melting temperature $T_m$ (and $F_0<0$ for $T<T_m$), while $\sigma_F$ is
proportional to the difference in melting temperatures for $A$ and $B$
homopolymers \cite{deG}.

Following the analogy with splitting probabilities, the probability that a
homopolymer (no random potential) ultimately melts when a length $Y$ is
initially in a random coil state is \cite{fpp}
\begin{equation}
\label{E+-}
E_+(Y)=\frac{1-e^{-vY/D}}{1-e^{-vL/D}}=\frac{1-z}{1-z^{1/f}}~,
\end{equation}
where, for later convenience, we define $z\equiv e^{-vY/D}$ and $f\equiv Y/L$.
Here $D$ is the diffusion coefficient of the boundary point between the melt
and helix (Fig.~\ref{cartoon}), and the velocity $v>0$ for $T>T_m$ and $v<0$
for $T<T_m$.  For $T =T_m$, corresponding to $v=0$, these splitting
probabilities simplify to $E_+(Y)=1-E_-(Y)=f$, the classical result for
unbiased diffusion \cite{fpp}.  We now focus on the corresponding behavior
for a heteropolymer at the melting temperature where the boundary point moves
in the random potential defined by Eq.~\eqref{force}.

\begin{figure}[ht]
\centerline{\includegraphics*[width=0.275\textwidth]{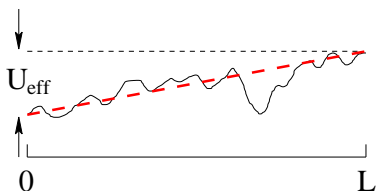}}
\caption{A typical realization of the potential in the continuum
  limit of the Sinai model (smooth curve) and the potential in the
  corresponding effective medium (dashed).}
  \label{EM}
\end{figure}

We first present an optimal fluctuation method \cite{lifshitz} to describe
the statistical features of $\mathcal{P}(E_+(Y))$ to describe the ultimate
fate of a single chain.  The basis of this approximation is to replace each
realization of the Sinai model by an effective environment whose mean bias
matches that of the Sinai model (Fig.~\ref{EM}).  Then the distribution of
bias velocities in the continuum limit is
\begin{equation}
\label{Pv}
P(v)= \frac{1}{\sqrt{2\pi\sigma_v^2}}\,\, e^{-v^2/2\sigma_v^2}~.
\end{equation}
Since there is a one-to-one connection between the bias in the effective
model and the splitting probability $E_+(Y)$, we may now convert the
distribution of bias velocities to the distribution of splitting
probabilities by
\begin{equation}
\label{tform}
\mathcal{P}(E_+)= P(v)\, \frac{dv}{d E_+}~.
\end{equation}

To perform this transformation, we use Eq.~\eqref{E+-} to solve for $v$ as a
function of $E_+$ for fixed $f$.  This inversion is feasible only for
$f=\frac{1}{2}, \frac{1}{3}, \frac{1}{4}, \frac{1}{5}$, because solving a
polynomial up to quartic order is involved.  For the simplest case of a
particle starting at the midpoint, $f=\frac{1}{2}$, then \eqref{E+-} becomes
$E_+=1/(1+z)$.  Inverting gives
\begin{equation}
\label{v}
v(E_+)=\frac{2D}{L}\ln\frac{E_+}{E_-}\qquad 
\frac{dv}{dE_+} =\frac{2D}{LE_+E_-}~.
\end{equation}
Substituting these results into Eq.~\eqref{Pv} and using \eqref{tform} gives,
\begin{eqnarray}
\label{PE}
\mathcal{P}(E_+)
\! =\! \frac{1}{\sqrt{2\pi \sigma_v^2}}
\frac{2D}{LE_+E_-}
\exp\left[-\frac{2D^2}{\sigma_v^2 L^2}
\ln ^2\!\left(\frac{E_+}{E_-}\right)\right]\!.
\end{eqnarray}

\begin{figure}[ht]
\centerline{\includegraphics*[width=0.405\textwidth]{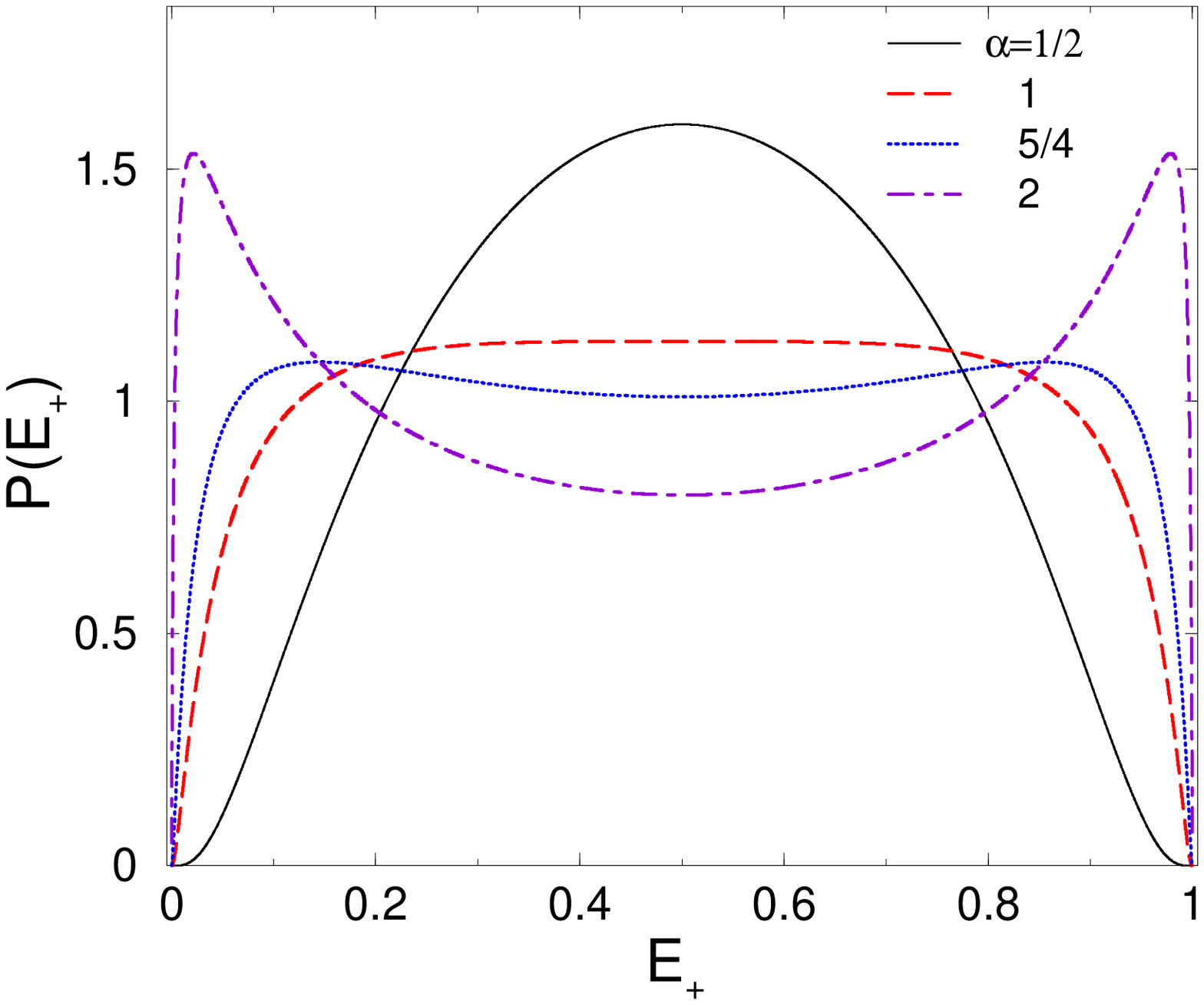}}
\centerline{\includegraphics*[width=0.425\textwidth]{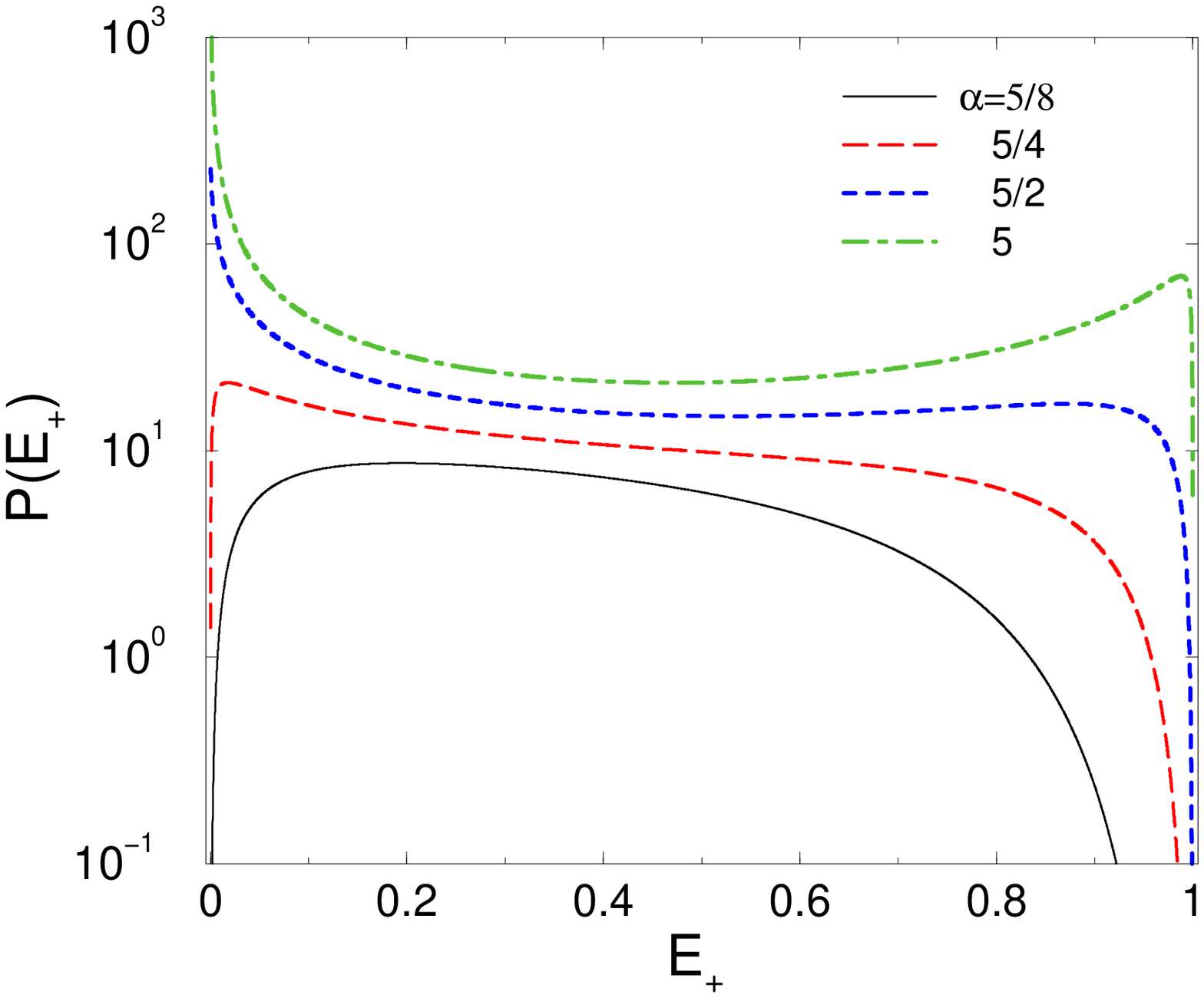}}
\caption{(color online) Optimal fluctuation prediction for the splitting probability
  distribution for the Sinai chain when the particle starts at $Y=L/2$ (top)
  and $Y=L/3$ (bottom, logarithmic scale).}
  \label{PE-fig}
\end{figure}

We now relate the velocity variance $\sigma_v^2$ to the disorder in the Sinai
model.  For Sinai disorder, the mean-square potential difference between two
points separated by a distance $L$ is (Fig.~\ref{EM})
\begin{equation*}
U_{\rm eff}^2\equiv\int_0^L\int_0^L\
\langle F(x')F(x'')\,\rangle dx'dx'' = \sigma^2_F \, L\,.
\end{equation*}
Thus there is a net force $F_{\rm eff}\sim U_{\rm eff}/L \sim
\sigma_F/\sqrt{L}$, from which we infer the velocity scale $\sigma_v\sim
\sigma_F/(\gamma \sqrt{L})$, where $\gamma$ is the viscosity coefficient.  We
then use the fluctuation-dissipation relation $D=kT/\gamma$
to rewrite Eq.~\eqref{PE} as
\begin{eqnarray}
\label{PE-F}
\mathcal{P}(E_+)\! =\! \frac{1}{\sqrt{4\pi \alpha}}
\frac{1}{E_+E_-} \exp\left[-\frac{1}{4\alpha}
\ln ^2\!\left(\!\frac{E_+}{E_-}\!\right)\right]\!,
\end{eqnarray}
with $\alpha = \sigma_F^2\,L/[8(kT)^2]$ a dimensionless measure of the
strength of the Sinai potential relative to thermal fluctuations.  The
important feature of this splitting probability distribution is its change
from unimodal to bimodal as the disorder parameter $\alpha$ increases past
$\alpha_c=1$ (Fig.~\ref{PE-fig}).  

In the critical case $\alpha = \alpha_c$ the maximum at $E_+ = 1/2$ is
quartic and the distribution is close to uniform for $0.2\alt E_+\alt 0.8$.
Thus any value of $E_+$ in this range is nearly equally probable.  For
$\alpha\agt\alpha_c$, two maxima emerge continuously at $E_+=\frac{1}{2}[1\pm
\sqrt{3(\alpha-\alpha_c)}]$.  Curiously, although the disorder-average value
$\langle E_{+}\rangle = 1/2$, for $\alpha > \alpha_c$ the probability
distribution has a minimum at $E_+ = 1/2$.  In this strong disorder limit,
the underlying mechanism for the bimodality of the distribution is that a
typical realization of the environment has a net bias (see Fig.~\ref{EM}) so
that the splitting probability is either close to zero or to one.

For $f=\frac{1}{3}$, a similar calculation gives the splitting probability
distribution shown in Fig.~\ref{PE-fig}.  For weak disorder, the splitting
probability is peaked at a point close to the value $\frac{1}{3}$ that arises
in the case of no disorder.  When $\alpha$ is sufficiently large, the
splitting probability has peaks near 0 and 1, corresponding to strong
disorder where a typical configuration has an overall net bias.  The
intermediate regime of $\alpha$ gives the strange situation in which there is
a peak at one extremum but not the other.

For a general starting point, the inversion $v=v(E_+)$ and thus the form of
$\mathcal{P}(E_+)$ can be determined asymptotically in the limits $|vL/D|\to
\infty$.  For $vL/D\to -\infty$, $z\to +\infty$, and Eq.~\eqref{E+-} becomes
$E_+\sim z^{-(1-f)/f}$.  Solving for $v(E_+)$ and $\frac{dv}{dE_+}$ and using
these results in Eq.~\eqref{tform}, the splitting probability distribution
starting from $Y=fL$ is
\begin{equation}
\label{E0}
\mathcal{P}(E_+) = \frac{1}{\sqrt{16\pi \alpha}} \frac{1}{(1\!-\!f)E_+}\,
\exp\left[-\frac{\ln^2 E_+}{16\alpha (1\!-\!f)^2}\right].
\end{equation}
As $\alpha$ is varied, this distribution changes from having a peak at the
most probable value of $E_+$ to a peak at $E_+=0$.  In the complementary
limit $vL/D\to +\infty$, corresponding to $z\to 0$ and $E_+\to 1$, the
distribution is again \eqref{E0}, but with $f\to 1-f$ and $E_+\to 1-E_+$.

We now give an exact solution for the splitting probabilities that
qualitatively substantiates the heuristic results given above.  Consider the
Fokker-Planck for the probability distribution for a particle that moves in
the Sinai potential, $P_t =D\left[P_{xx}- \beta (F(x) P)_x\right]$, where
$\beta= 1/kT$ and $P(x,t)\,dx$ is the probability that the particle is in the
range $[x,x+dx]$ at time $t$ and $F(x)$ is the force at $x$.  Suppose that
particles are injected at a constant rate at $Y$ to maintain a fixed
concentration $P(Y)=1$ when absorbing boundary conditions at $x=0$ and $x=L$
are imposed.  Then the stationary solution to the Fokker-Planck equation is
\begin{equation}
P(x) =
\begin{cases}
e^{\beta\mathcal{F}_Y(x)}
 \left[1 -{\displaystyle \frac{\int_Y^x  \exp[-\beta\mathcal{F}_Y(x)]\, dx}
{\int_Y^L  \exp[-\beta\mathcal{F}_Y(x)]\, dx}} \right] & x \geq Y \\ \\
e^{\beta\mathcal{F}_Y(x)}
 \left[1 - {\displaystyle\frac{\int^{x}_{Y}  \exp[-\beta\mathcal{F}_Y(x)] dx}
{\int^{0}_{Y}  \exp[-\beta\mathcal{F}_Y(x)]\, dx}} \right] & x \leq Y,
\end{cases}
\end{equation}
where $\mathcal{F}_Y(x)\equiv \int^{x}_{Y} dx' F(x')$.

From the steady current $J=-D[P_x-\beta(F(x)P)]$, the splitting
probabilities $E_\pm(Y)$ are simply
\begin{eqnarray}
E_- = \frac{\tau_+}{\tau_- + \tau_+}~, \qquad
E_+ = \frac{\tau_-}{\tau_- + \tau_+}
\end{eqnarray}
where $\tau_\pm=1/J_\pm$, with $J_\pm$ being the flux to the boundaries at 0
and $L$, respectively.  Thus $\tau_\pm$ define the ``resistance'' of a finite
interval with respect to passage across the two boundaries and are given
explicitly by
\begin{align}
\label{tau}
\tau_- = \int_0^Y  e^{-\beta \mathcal{F}_0(x)}\, dx, \quad
\tau_+ = \int_0^{L-Y} e^{-\beta \mathcal{F}_0(x)}\, dx.
\end{align}
We emphasize that the force integrals $\mathcal{F}_0$ are {\em independent\/}
variables in each integrand so that $\tau_-$ and $\tau_+$ are also {\em
  independent\/} random variables.

The quantities defined in \eqref{tau} are the continuous-space counterparts
of Kesten variables \cite{kesten} that play an important role in many
stochastic processes.  For example, negative moments of $\tau$ describe the
positive moments of a steady diffusive current in a finite Sinai chain
\cite{bur1,osh1,osh2,comtet1,comtet2}.  A surprising feature is that the
disorder-average currents $J_\pm$ scale with system length $L$ as $L^{-1/2}$
and are much larger than Fickian which scale as $L^{-1}$ in homogeneous
environments.  Thus Sinai chains have an anomalously high conductance despite
the fact that diffusion is logarithmically confined \cite{sinai} and positive
moments of $\tau$ grow as $e^{\sqrt{L}}$
\cite{bur1,osh1,osh2,comtet1,comtet2}.

Let $\Psi_-(\tau_-)$ and $\Psi_+(\tau_+)$ denote the distribution functions
of the random variables $\tau_-$ and $\tau_+$, respectively.  Then the moment
generating function of the splitting probability $E_+$ can be written as:
\begin{equation}
\langle e^{-\lambda E_+} \rangle\! =\! \int_0^{\infty} \!\!\!\!\int_0^{\infty}
\!\!\!\Psi_-(\tau_-) \Psi_+(\tau_+)\, e^{-\lambda \tau_-/(\tau_- + \tau_+)}\,  d\tau_- d\tau_+.
\end{equation}
Integrating over $d\tau_+$, we formally change the integration variable from
$\tau_+$ to $E_+$ to give
\begin{eqnarray*}
\langle e^{-\lambda E_+} \rangle =
\int^{1}_0 \frac{d E_+}{E_+^2} e^{-\lambda E_+} 
\!\!\! \int^{\infty}_0 \!\!\!\!d\tau_-\, \tau_- \Psi_-(\tau_-)
\Psi_+(r \tau_-),
\end{eqnarray*}
where $r=E_-/E_+$.  From this expression we may read off the following form
of the probability density $\mathcal{P}(E_+) $ of the splitting probability
$E_+$:
\begin{equation}
\label{dist}
\mathcal{P}(E_+)  = \frac{1}{E_+^2} \int^{\infty}_0 d\tau_-\, \tau_- \Psi_-(\tau_-) 
\Psi_+(r \tau_-).
\end{equation}

For a random potential without any bias, the distribution $\Psi_-(\tau_-)$ is
given by (see, e.g., Ref.~\cite{comtet2}):
\begin{eqnarray}
\label{psi-}
\Psi_-(\tau_-) \!\!&=&\!\! \frac{\sqrt{fL}}{2\pi\alpha_- \tau_-^{3/2}} 
\int_{-\infty}^{\infty}\!\!\! du\, {\rm ch}\, u\, 
\cos\left(\frac{\pi u}{2\alpha_-}\right) \nonumber\\
&\times& \exp\left[\frac{\pi^2}{8 \alpha_-} - \frac{u^2}{2\alpha_-} 
- \frac{ 2 fL\,\,{\rm ch}^2 u}{4\alpha_- \tau_-}\right],
\end{eqnarray}
where $\alpha_- = \beta^2 \sigma_F^2 fL/4$ and ${\rm ch}\, u$ denotes the
hyperbolic cosine.  Then $\Psi_+(\tau_+)$ is obtained from \eqref{psi-} by
replacing $\tau_- \to \tau_+$ and $\alpha_- \to \alpha_+ = \beta^2 \sigma_F^2
(1-f)L/4$.

\begin{figure}[ht]
\centerline{\includegraphics*[width=0.425\textwidth]{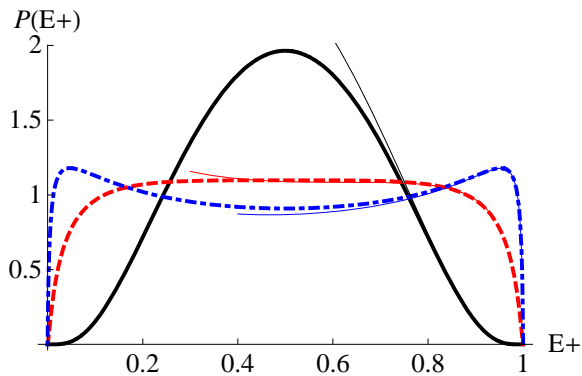}}
\caption{(color online) Exact distribution $\mathcal{P}(E_+)$, Eq.~(\ref{dist3}).  
  for $\alpha_-=\alpha_+=1/4$ (solid curve) $\alpha_\pm=0.815$ (dashed), and
  $\alpha_\pm=1.2$ (dash-dotted).  Thin curves are the corresponding
  asymptotic results of Eq.~(\ref{as1}). }
\label{PE-exact}
\end{figure}

Substituting $\Psi_-(\tau_-)$ and $\Psi_+(\tau_+)$ into Eq.~(\ref{dist}) and
integrating over $d \tau_-$, we find 
\begin{eqnarray}
\label{dist2}
\mathcal{P}(E_+) &=& \frac{1}{\pi^2 \sqrt{4\alpha_- \alpha_+ E_+ E_-}} 
\int_{-\infty}^{\infty}\!\int_{-\infty}^{\infty}  du_1\, du_2  \nonumber\\
&\times&    \frac{\,{\rm ch}\, u_1\, {\rm ch}\, u_2}{E_- {\rm ch}^2 u_1 + E_+ {\rm ch}^2 u_2} 
\cos\left(\frac{\pi u_1}{2\alpha_-}\right) \cos\left(\frac{\pi u_2}{2\alpha_+}\right) \nonumber\\
&\times&   \exp\left[\frac{\pi^2}{8 \alpha_-} + \frac{\pi^2}{8 \alpha_+} - 
\frac{u_1^2}{2\alpha_-} - \frac{u_2^2}{2\alpha_+}\right].
\end{eqnarray}
After cumbersome but straightforward manipulations, one of the integrals can
be performed to recast the splitting probability distribution as:
\begin{eqnarray}
\label{dist3}
&\mathcal{P}(E_+)& =  \frac{1}{\pi \sqrt{4\alpha_- \alpha_+ E_+ E_-}} \frac{1}{E_-} 
 \int_{-\infty}^{\infty} du\, \frac{{\rm ch}\, u}{{\rm ch}\,\eta} 
\cos\left(\frac{\pi u}{2\alpha_+}\right) \nonumber\\
&\times&\exp\left[\frac{\pi^2}{8 \alpha_+} - \frac{u^2}{2\alpha_+} -
\frac{\eta^2}{2\alpha_-}\right], 
\end{eqnarray}
where $\eta = {\rm sh}^{-1}\left[({\rm ch}\, u)/\sqrt{r}\right]$.  For the
symmetric initial starting point, $\alpha_+=\alpha_-=\alpha$, the
distribution changes from unimodal to bimodal as $\alpha$ increases beyond
$\alpha_c\approx 0.815$ (Fig.~\ref{PE-exact}).

To extract the asymptotic behavior of $\mathcal{P}(E_+)$ in the limits
$E_+\to 1$ and $E_+\to 0$ note that because the integrand in
Eq.~(\ref{dist3}) contains an oscillating cosine term, only the behavior near
$u = 0$ should matter.  Thus assuming ${\rm ch}\, u \approx 1$ and expanding
$\eta^2$ in a Taylor series in $u$ up to second order,
we find the following asymptotic representation for $E_+\to 1$:
\begin{eqnarray}
\label{as1}
&\mathcal{P}(E_+) & \sim \frac{\left(\alpha_- +\alpha_+ \sqrt{E_+} 
\ln z_+\right)^{-1/2}}{\sqrt{2\pi E_+}\,\, E_-}  \nonumber\\
&\times&\!\!\!\!\!\!\!\!\! \exp\!\left[- \frac{\ln^2 z_+}{2\alpha_-}
\! +\! \frac{\pi^2 \sqrt{E_+} \ln z_+}
{8\left( \alpha_- +\alpha_+ \sqrt{E_+} \ln z_+\right)}\right]\!,
\end{eqnarray}
where $z_+ = (1 + \sqrt{E_+})/\sqrt{E_-}$.  This asymptotic form agrees quite
well with the exact result in Eq.~(\ref{dist3}), not only when $E_+\to 1$,
but also for moderate values of $E_+$ (Fig.~\ref{PE-exact}).  Similarly, we
obtain the asymptotics of $\mathcal{P}(E_+)$ for $E_+\to 0$ merely by
interchanging all subscripts $\pm$ to $\mp$ in \eqref{as1}, and $z_-$ is
obtained from $z_+$ by interchanging the subscripts $\pm$ to $\mp$ in the
latter.  When $E_+ \ll 1$ or $1 - E_+ \ll 1$ respectively, Eq.~(\ref{as1})
reduces to
\begin{eqnarray}
\mathcal{P}(E_+) \sim \sqrt{\frac{1}{\pi \alpha_\pm  \ln\left(1/E_\mp\right)  }} 
\frac{1}{E_\mp } 
e^{- \frac{1}{8 \alpha_\mp} \ln^2\left(\frac{1}{E_\mp}\right)}~.
\end{eqnarray}
Our optimal fluctuation result \eqref{PE-F} resembles the above form apart
from logarithmic and numerical factors.

In summary, by making an analogy to first passage on a Sinai chain, we find
that the evolution of a partially melted random heteropolymer at the melting
temperature is controlled by the sequencing of monomers.  Each heteropolymer
realization has a unique kinetics and final fate that is not representative
of the average behavior of an ensemble of such polymers.  A related lack of
self averaging was recently found in anomalous diffusion \cite{LSK}.

{\bf Acknowledgments.} We gratefully acknowledge helpful discussions with O.
B\'enichou.  GO is partially supported by Agence Nationale de la Recherche
(ANR) under grant ``DYOPTRI - Dynamique et Optimisation des Processus de
Transport Intermittents'' while SR is partially supported by NSF grant
DMR0535503 and the University of Paris VI.

 \end{document}